\documentclass[aps,prb,letterpaper,amsmath,amssymb,reprint, superscriptaddress]{revtex4-1}
\usepackage{siunitx}
\usepackage{graphicx}
\usepackage{microtype}
\usepackage{bm}
\usepackage{hyperref}
\usepackage{color}

\usepackage{amsmath}

\usepackage{ulem}

\newcommand{\tn}{\textnormal}

\newcommand{\I}{\tn{i}} 
\newcommand{\fig}{Fig.\,\ref}
\newcommand{\figs}{Figs.\,\ref}

\newcommand{\figus}{Figures\,\ref}

\newcommand{\eq}{Eq.\,\eqref}
\newcommand{\Eq}{Equation\,\eqref}

\newcommand{\Refs}{Refs.\,\onlinecite}
\newcommand{\E}{\tn{e}}

\begin{document}
\title{High-throughput techniques for measuring the spin Hall effect}
\author{Markus Meinert}\email{markus.meinert@tu-darmstadt.de}\affiliation{Department of Electrical Engineering and Information Technology, Technical University of Darmstadt, Merckstraße 25, D-64283 Darmstadt, Germany}

\author{Bj\"orn Gliniors}\affiliation{Center for Spinelectronic Materials and Devices, Department of Physics, Bielefeld University, D-33501 Bielefeld, Germany}

\author{Oliver Gueckstock}\affiliation{Department of Physics, Freie Universit\"at Berlin, Berlin, Germany}\affiliation{Fritz Haber Institute of the Max Planck Society, Berlin, Germany}

\author{Tom S. Seifert}\affiliation{Department of Physics, Freie Universit\"at Berlin, Berlin, Germany}\affiliation{Fritz Haber Institute of the Max Planck Society, Berlin, Germany}

\author{Lukas Liensberger}
\affiliation{Walther-Mei{\ss}ner-Institut, Bayerische Akademie der Wissenschaften, Garching, Germany}\affiliation{Physik-Department, Technische Universit\"at M\"unchen, Garching, Germany}

\author{Mathias Weiler}
\affiliation{Walther-Mei{\ss}ner-Institut, Bayerische Akademie der Wissenschaften, Garching, Germany}\affiliation{Physik-Department, Technische Universit\"at M\"unchen, Garching, Germany}
\affiliation{Fachbereich Physik, TU Kaiserslautern, Kaiserslautern, Germany}

\author{Sebastian Wimmer}\affiliation{Department Physik, Ludwig-Maximilians-Universit\"at M\"unchen, M\"unchen, Germany}
\author{Hubert Ebert}\affiliation{Department Physik, Ludwig-Maximilians-Universit\"at M\"unchen, M\"unchen, Germany}

\author{Tobias Kampfrath}\affiliation{Department of Physics, Freie Universit\"at Berlin, Berlin, Germany}\affiliation{Fritz Haber Institute of the Max Planck Society, Berlin, Germany}
\date{\today}

\begin{abstract}
The spin Hall effect in heavy-metal thin films is routinely employed to convert charge currents
into transverse spin currents and can be used to exert torque on adjacent ferromagnets. Conversely,
the inverse spin Hall effect is frequently used to detect spin currents by charge currents in
spintronic devices up to the terahertz frequency range. Numerous techniques to measure the spin
Hall effect or its inverse were introduced, most of which require extensive sample preparation by
multi-step lithography. To enable rapid screening of materials in terms of charge-to-spin
conversion, suitable high-throughput methods for measuring the spin Hall angle are required. Here,
we compare two lithography-free techniques, terahertz emission spectroscopy and broadband
ferromagnetic resonance, to standard harmonic Hall measurements and theoretical predictions using
the binary-alloy series Au$_x$Pt$_{1-x}$ as benchmark system. Despite being highly complementary,
we find that all three techniques yield a spin Hall angle with approximately the same
$x$~dependence, which is also consistent with first-principles calculations. Quantitative
discrepancies are discussed in terms of magnetization orientation and interfacial spin-memory
loss.\end{abstract}

\maketitle

\begin{figure*}[th!]
\includegraphics[width=\textwidth]{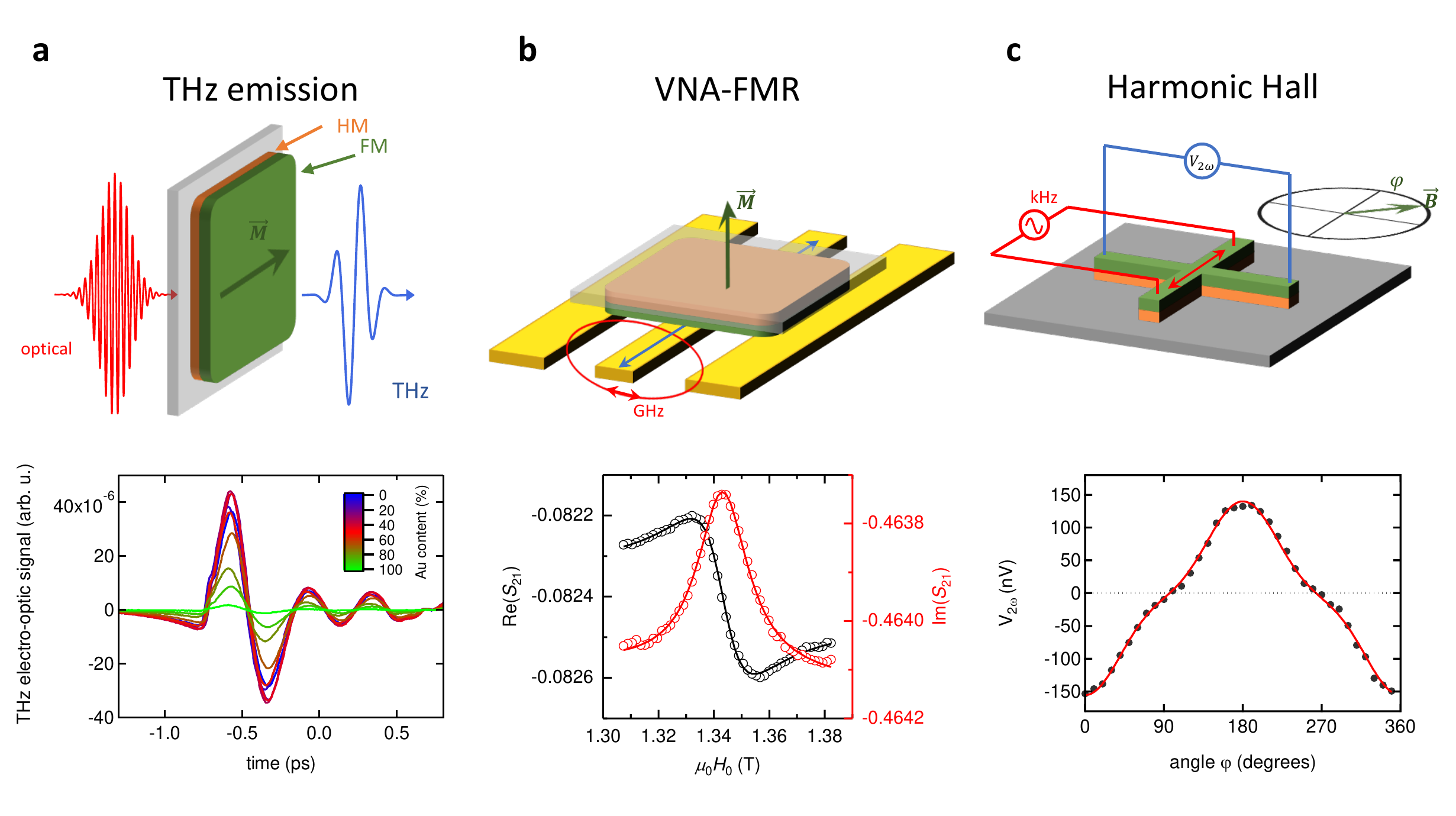}
\caption{\label{schematics} Overview of three different techniques to determine the spin Hall angle
of a material as described in the text. The top row shows schematic of the techniques, while the
bottom row shows typical raw data. (a)~THz emission spectroscopy: an optical laser pulse generates an ultrafast heat pulse in the films. Due to the spin-dependent Seebeck effect, a spin current flows from the ferromagnetic layer (FM) into the heavy metal layer (HM). The inverse spin Hall effect converts the spin current into a charge pulse, which emits THz radiation. (b)~Vector-network-analyzer
ferromagnetic resonance (VNA-FMR): a GHz current in the coplanar waveguide excites the ferromagnetic resonance in the FM. Spin pumping drives a spin current into the HM layer, where it is converted into an oscillating charge current. Its magnetic field couples into the waveguide and can be detected in the complex-valued waveguide transmission signal $S_{21}$.  (c)~Harmonic Hall measurements: A kHz charge current drives an oscillating spin current from the HM into the FM layer. The associated spin-orbit torque drives an oscillating deflection of the magnetization out of the film plane. The associated oscillating anomalous Hall voltage is detected as a second-harmonic transverse voltage in the Hall cross.}
\end{figure*}

\section*{Introduction}

The spin Hall effect\cite{Dyakonov1971, Hirsch1999, Hoffmann2013, Sinova2015} (SHE) converts a
charge current with density $j_\tn{c}$ into a transverse spin current with density $j_\tn{s}$. The
charge-to-spin conversion efficiency can be characterized by the spin Hall angle (SHA)
$\theta_\tn{SH} = j_\tn{s}/j_\tn{c}$. It is commonly written as $\theta_\tn{SH} =
\sigma_\tn{SH}/\sigma_{xx}$, where $\sigma_\tn{SH}$ is the spin Hall conductivity (SHC) and
$\sigma_{xx}$ is the longitudinal conductivity along the direction of the charge-current flow. Both
intrinsic effects that are already present in perfectly periodic crystals and extrinsic effects,
i.e. skew scattering and side-jump scattering, contribute to the spin Hall angle.
In most practical cases, the electron scattering rate in a material is large due to
point defects, grain boundaries, and phonons; therefore, the intrinsic mechanism dominates
$\sigma_\tn{SH}$. Following theoretical predictions,\cite{Tanaka2008} various crystalline heavy
metals (HM) with large spin Hall conductivity were experimentally confirmed, such as
Pt\cite{Sagasta2016}, $\beta$-W \cite{Pai2012} and $\beta$-Ta \cite{Liu2012a}.

The SHC of crystalline materials is experimentally\cite{Qiu2013, Sagasta2016, Nguyen2016,
Schulz2016, Zhang2015} and theoretically\cite{Tanaka2008, Lowitzer2011, Freimuth2010, Gradhand2012,
Koedderitzsch2015} well understood. The relation $\theta_\tn{SH} = \sigma_\tn{SH}/\sigma_{xx}$ was
experimentally studied for Pt thin films\cite{Sagasta2016}. This study partially explains the large
range of reported SHAs in the literature for a single material and rationalizes the somewhat
counterintuitive observation that thin films of lower quality and, thus, lower conductivity
$\sigma_{xx}$ have a larger SHA\cite{Sinova2015}. By alloying Pt with Au, it was
shown\cite{Obstbaum2016} that $\sigma_{xx}$ can be decreased to increase the SHA.

The spin current originating from the SHE can be injected into an adjacent ferromagnetic (FM) layer
where it gives rise to so-called field-like and damping-like spin-orbit torques \cite{Manchon2009,
Garello2013}. These may induce precession of the magnetization \cite{Liu2012b}, domain-wall motion
\cite{Miron2010} or switching of the magnetization orientation\cite{Miron2011, Pai2012, Liu2012a}.
Various promising concepts for SHE-based magnetic memory devices, so called spin-orbit torque
magnetic random-access memories (SOT-MRAMs) were proposed\cite{Cubukcu2014, Garello2014,
Fukami2016,
Lau2016}. 

To quantify the SHA or the SHC experimentally, numerous techniques were developed, as detailed in
the reviews of \Refs{Hoffmann2013, Sinova2015} and references therein. The various techniques can
differ significantly, for example with respect to the driving perturbation, probed observable,
magnetization and external-field geometry, use of either the direct or the inverse SHE and covered
frequency range. On one hand, several ferromagnetic-resonance (FMR)-based techniques were used,
such as (i)~FMR spin pumping with subsequent detection of the inverse SHE (ISHE)\cite{Weiler2014}
and (ii)~SHE-induced modulation of the linewidth in an FMR experiment due to the action of the
damping-like torque from the spin current\cite{Liu2011}. The quantification of the SHE relies on
either measuring the rectified voltages generated by the inverse SHE, or on determining the
modulation of the FMR linewidth, which originates from the damping-like spin-orbit torque. On the
other hand, electrical transport techniques were developed that employ nonlocal spin
injection,\cite{Kimura2007} the spin Hall magnetoresistance,\cite{Nakayama2013} magnetic loop
shifts\cite{Pai2016} or measure the deflection of the magnetization and the resulting change in the
anomalous Hall effect.\cite{Liu2012a} All of these techniques are in principle quantitative and
have in common that relatively tedious lithographic preparation of microdevices is required.

Materials with large SHA can be tailored by alloying,\cite{Obstbaum2016, Qu2018, Derunova2018},
which decreases the conductivity and tunes the Fermi level close to maxima of the SHC. Also, phase
transitions in binary or ternary phase diagrams can be exploited, or amorphous metals with low
conductivity may be created by enforced mixing of immiscible elements.\cite{Fritz2018} The
associated maxima in the SHA as a function of composition can be quite narrow, such
that optimization requires a large number of samples to be investigated. Lithography is
time-consuming and, thus, a limiting factor for high-throughput SHA characterization. Therefore,
compatible methods that do not require any additional processing steps are highly desirable.
Recently, two promising techniques potentially fulfilling this need have become available: THz
emission spectroscopy (TES, \fig{schematics}a, top) and vector network analyzer (VNA) FMR (VNA-FMR,
\fig{schematics}b, top). Despite their relevance for rapid sample characterization, their
performance has not yet been compared to each other and a well-established technique such as
harmonic Hall response (HHR, \fig{schematics}c, top). Recent work that experimentally compared the
spin Seebeck effect at dc and THz frequencies indicates that TES has large potential for material
characterization with results that are consistent with static methods \cite{Cramer2018}.


In this article, we demonstrate that both TES and VNA-FMR are suitable techniques to quickly obtain
quantitative measurements of the SHA of a metallic binary alloy series. Both methods do not require any
post-deposition sample processing and feature data acquisition times on the timescale of minutes (TES) to
hours (VNA-FMR). We compare the results from these two high-throughput methods to harmonic Hall response
measurements, which utilize a single lithography step and serve as a reference.\cite{Pi2010,
Hayashi2014, Avci2014, Wen2017} In addition, we compare our results with first-principles
calculations.\cite{Obstbaum2016} Although the three methods are very different in terms of
frequency windows, spin current generation, and detection schemes, we find a qualitatively good
agreement between them.

\section*{Experimental details}

\subsection*{Sample system}

To benchmark our techniques, we use a series of Au-Pt binary alloys. Thin-film stacks of
Au$_x$Pt$_{1-x}$(3~nm)$|$CoFeB(3~nm)$|$Si(1.5~nm) were grown by dc unbalanced magnetron
co-sputtering in a 2" sputtering system at room temperature. For TES and VNA-FMR experiments, the
samples were not processed any further. For the harmonic Hall measurements, the samples were
patterned with Hall cross devices with fourfold rotational symmetry and a line width of 16\,$\mu$m
by standard optical lithography and Ar-ion-beam milling. All samples were checked by X-ray
diffraction, X-ray reflectivity, X-ray fluorescence and four-point dc conductivity measurements.

The in-plane conductivities $\sigma_{xx}$ of the Au-Pt alloy layers are shown in
\fig{spinhallangles}a, where a parallel-conductor model was applied to remove the contributions
from the CoFeB layers ($\sigma_\mathrm{CoFeB} = 5.7 \times 10^{5}$\,S/m). The corresponding 
conductivity of the Pt layer ($2.78 \times
10^{6}$\,S/m) is quite typical for a thickness of 3~nm.\cite{Sagasta2016, Nguyen2016, Dutta2017}
As expected, doping with Au reduces $\sigma_{xx}$ substantially. The conductivity of Au-rich
samples remains low, because of the pronounced island growth of Au on SiO$_2$ surfaces. In the
following, we discuss key aspects of the employed spin Hall measurement techniques and present
respective results. Additional technical details are provided in the Appendix.

\subsection*{Method~(a): Terahertz emission spectroscopy}

In TES, a HM$|$FM bilayer under study is excited by a femtosecond laser pulse (\fig{schematics}a,
top), thereby inducing ultrafast spin transport from the FM into the HM layer through an ultrafast
version of the spin-dependent Seebeck effect.\cite{Seifert2016, Kampfrath2013, Seifert2017,
Alekhin2017} In the HM, the laser-driven longitudinal spin current is converted into a transverse
charge current by the ISHE. The resulting sub-picosecond charge current\cite{Seifert2018} gives
rise to the emission of electromagnetic radiation at THz frequencies.\cite{Seifert2016}

The THz waveforms of \fig{schematics}a (bottom) are raw data obtained with this technique. The
emission amplitudes are modeled as a function of THz frequency $\omega/2\pi$ as\cite{Seifert2016}
\begin{equation}\label{eq:THz}
S_\tn{THz}(\omega) = {AB(\omega)\lambda_\tn{s}}\cdot\tanh\left(\frac{t_\tn{HM}}{2\lambda_\tn{s}}\right)\cdot\theta_\tn{SH}\cdot Z(\omega).
\end{equation}
Here, $A$ is the pump-light absorptance, while the factor $B$ captures the
photon-to-spin-current conversion efficiency and the detector response function\cite{Braun2016}.
$B$~is assumed to be independent of the alloy composition in our experiment, thereby 
neglecting possible variations of the spin-current
strength due to, e.g., variations of the interface quality for different Au concentrations.

The spatial shape of the spin current in the HM layer is captured by the spin-current relaxation
length $\lambda_\tn{s}$ and the HM layer thickness $t_\tn{HM}$. According to transport theory based
on the Boltzmann equation\cite{HCSchneider2008}, $\lambda_\tn{s}$ equals the spin diffusion length
at zero frequency, but becomes comparable to the mean free path length at THz frequencies.

While the spin-to-charge-current conversion in \eq{eq:THz} is quantified by $\theta_\tn{SH}$, the
charge-current-to-electric-field conversion is described by the bilayer impedance
\begin{equation}\label{eq:impedance}
Z(\omega) = \frac{Z_0}{n_1(\omega) + n_2(\omega) + Z_0 \int_0^d \tn{d}z \sigma_{xx}(z, \omega)}
\end{equation}
where $n_1(\omega)$ and $n_2(\omega)$ are the refractive indices of air and the substrate,
respectively, $Z_0 = 377\,\Omega$, and $\sigma_{xx}(z, \omega)$ is the in-plane conductivity of the
material at depth $z$. For simplicity, we take $\sigma_{xx}$ as constant across the film thickness
and ignore the frequency dependence, because the frequencies used here are well below the Drude
frequency of the material.

The SHA relative to a reference sample can be obtained for all alloy stoichiometries when
$\lambda_\tn{s}$ is known. Here, we take $\lambda_\tn{s}$ as the electron mean free path
$\lambda_\tn{MF}$ and use $\lambda_\tn{MF}/\sigma_{zz} = 0.3\times 10^{-15}~\Omega~\tn{m}^2$ where
$\sigma_{zz}$ is the electrical conductivity of the HM perpendicular to the film
plane.\cite{Sagasta2016, Dutta2017} Measuring $\sigma_{zz}$ is impractical, so we employ the
approximation $\sigma_{zz} \approx \sigma_{xx}$.

Note that $\theta_\tn{SH}$ of \eq{eq:THz} is an effective SHA which, in addition to
spin-to-charge-current conversion in the HM layer, contains such conversion also in the FM layer
and at the FM/HM interface. Notably, all THz measurements, i.e. THz emission, pump absorptance and
THz conductivity, were conducted within less than 8~h.

\subsection*{Method~(b): VNA ferromagnetic resonance}

In VNA-FMR, we inductively detect microwave currents generated in HM$|$FM bilayers under the
condition of FMR, which allows one to determine the SHC. The sample is placed face-down on a
coplanar waveguide (CPW) (\fig{schematics}b, top). A GHz current excites resonant spin precession
(FMR) in the FM part of the bilayer. Due to spin pumping, a spin current flows into the HM layer
where it is converted into a charge current by the ISHE. The magnetic field created by this current
couples back into the CPW and is extracted from the CPW transmission signal to obtain the
complex-valued  SOT conductivity $\sigma^\tn{SOT}$. This quantity is directly linked to the SHC.\cite{Berger2018_1, Berger2018_2}

Raw data obtained by VNA-FMR are the real and imaginary part of the CPW transmission $S_{21}$ as a function of external magnetic field at fixed continuous-wave frequency (\fig{schematics}b, bottom). The $S_{21}$ data is fitted to Eq.~\eqref{eq:S21fitting} and from \eq{eq:Ltilde} (see Methods section),
we obtain the complex-valued normalized inductance $\widetilde{L}$ of the HM$|$FM bilayer at frequency $\omega$. For each sample, the $S_{21}$ measurements and extraction of $\widetilde{L}$ are performed for frequencies $5~\tn{GHz} < \omega/2 \pi < 40~\tn{GHz}$. The generation of charge currents in the
HM$|$FM bilayer under FMR conditions results in a linear frequency dependence of $\widetilde{L}$.
The dc value $\widetilde{L}(\omega=0)$ is the real-valued inductance of the HM$|$FM bilayer in the
absence of any currents in the bilayer. To extract the complex-valued SOT conductivity
$\sigma^\tn{SOT}=\sigma_\tn{e}^\tn{SOT}+\I\sigma_\tn{o}^\tn{SOT}$, $\widetilde{L}$ is fitted by
\cite{Berger2018_1}
\begin{equation}\label{eq:Lfreq}
\widetilde{L}\E^{\I\phi_\tn{a}}= \eta^2 \frac{\mu_0 t_\tn{FM} l}{4 w_\tn{C}}+\eta\omega\frac{\hbar L_{12}\sigma^\tn{SOT}}{2eM_\tn{s}}.
\end{equation}
The first term on the right side of~\eqref{eq:Lfreq} is the frequency-independent dipolar inductance stemming from the precessing magnetization. The second term is the linearly frequency-dependent inductance due to the ac currents flowing in the normal metal~\cite{Berger2018_1}. In~\eqref{eq:Lfreq},
$w_\tn{C}=56~\mu\tn{m}$ is the width of the CPW center conductor, $l=8.7~\tn{mm}$ is the sample
length, $L_{12}(d)$ is the mutual inductance between sample and CPW and $0<\eta(d)<1$ is a unitless
spacing loss as defined in~\cite{Berger2018_1}. Fit parameters are the separation $d$ between sample
and CPW, the anomalous phase $\phi_\tn{a}$ and the spin-orbit torque conductivities
$\sigma_\tn{e}^\tn{SOT}$ and $\sigma_\tn{o}^\tn{SOT}$, where the even component $\sigma_\tn{e}^\tn{SOT}$ also contains
the effect of currents induced by Faraday's law of induction. The odd component
$\sigma_\tn{o}^\tn{SOT}$ is directly related to the damping-like spin-orbit torque.

While the SOT conductivities can thus directly be measured using VNA-FMR, extraction of the microscopic
parameters, in particular the spin Hall angle, requires use of a suitable model and parameters~\cite{Berger2018_1, Berger2018_2}.
A lower limit of the spin Hall angle can be obtained by
\begin{equation}\label{eq:thetaSH_VNAFMR}
\theta_\tn{SH}=\sigma_\tn{o}^\tn{SOT}/\sigma_{xx}\,,
\end{equation}
where we again assume $\sigma_{xx} \approx \sigma_{zz}$. We note that Eq.~\eqref{eq:thetaSH_VNAFMR} assumes a completely transparent interface and thus vanishing spin backflow and spin memory loss (SML). As previously demonstrated~\cite{Berger2018_1, Berger2018_2}, we thus may underestimate the spin Hall angle by a factor $\approx10$. This underestimation is predominantly caused by the expected strong SML in HM/FM bilayers~\cite{Berger2018_1, Berger2018_2}. Quantification of the SML is in principle possible, but would require a thickness-series of both HM and FM layers for each composition~\cite{Berger2018_2}.  We use Eq.~\eqref{eq:thetaSH_VNAFMR} here for a fair comparison of the VNA-FMR evaluation to the reference measurements based on the Harmonic Hall response (see next section), where the same assumptions are made.

\subsection*{Method~(c): Harmonic Hall response}

Harmonic Hall voltage measurements are performed by injecting an ac current with amplitude $I_0$ at
frequency $\omega/2\pi$ into the Hall crosses measuring the in-phase first harmonic and
out-of-phase second harmonic Hall voltages simultaneously upon in-plane field rotation with a
lock-in amplifier (\fig{schematics}c, top). The SOT gives rise to a periodic deflection of the magnetization 
with in-plane and out-of-plane components, which can be detected via the planar Hall effect and the
anomalous Hall effect, respectively. The second-harmonic out-of-phase Hall voltage rms value
$V_{2\omega}$ depends on the in-plane angle $\varphi$ between current and magnetization
(\fig{schematics}c, bottom) and can be written as\cite{Wen2017,Fritz2018}
\begin{equation}\label{eq:harmonichall}
V_{2\omega} =  \left(-\frac{B_\tn{FL}}{B_\tn{ext}} R_\tn{P}\cos(2\varphi)
 - \frac{1}{2} \frac{B_\tn{DL}}{B_\tn{eff}} R_\tn{A} + \alpha' I_0 \right) I_\tn{rms}\cos\varphi.
\end{equation}
Here, $B_\tn{eff} = B_\tn{ext} + B_\tn{sat}$ is the effective field, $B_\tn{FL}$ and $B_\tn{DL}$
are the current-induced effective field amplitudes associated with the field-like (FL) and
damping-like (DL) spin-orbit torques.\cite{Garello2013} It is assumed that in-plane anisotropy
fields (e.g. uniaxial and biaxial) are small compared to the external magnetic field $B_\tn{ext}$
(0.2~T to 1~T) and can be neglected. The term $\alpha' I_0$ describes a parasitic contribution
arising from the anomalous Nernst effect (ANE),\cite{Avci2014}where $I_\mathrm{rms} = I_0 /
\sqrt{2}$. $R_\mathrm{P}$ and $R_\mathrm{A}$ are the amplitudes of the planar and anomalous Hall
resistances at saturation, respectively. $B_\mathrm{ext}$ is the external magnetic field and
$B_\mathrm{sat}$ is the perpendicular saturation field.

\Eq{eq:harmonichall} is fitted to the experimental data, and damping-like effective fields and
anomalous-Nernst contributions are separated by their dependence on the external field. The spin
Hall angle is obtained from the damping-like effective field as
\begin{equation}\label{eq:harmonichall_SHA}
\theta_\tn{SH} = \frac{2e}{\hbar} \frac{ B_\tn{DL} M_\tn{s} t_\tn{FM} }{j_{\tn{HM}0}},
\end{equation}
where $j_{\tn{HM}0}$ is the current density amplitude in the heavy-metal layer far away from the
Hall voltage pickup lines. In this expression, effects of spin memory loss, spin backflow or spin
transparency of the interface are neglected. Therefore, the SHA obtained by this formula is a
strict lower bound to the true SHA of the HM layer. A correction factor of 1.45 for the
inhomogeneous current flow in the Hall crosses was applied to the spin Hall angle, as suggested by
a recent study on the influence of the aspect ratio of the Hall cross on the effective field
determination.\cite{Neumann2018}

\subsection*{First-principles calculations}

For the first-principles calculations, we employ the Kubo-Bastin linear response theory as
implemented in the Munich SPR-KKR package.\cite{Ebert2015, Ebert2011, SPRKKR} Starting from a
density functional theory description of the electronic structure of the chemically disordered
alloy, linear response calculations including phonon effects via the alloy analogy model are
performed to obtain the full spin-resolved conductivity tensor. The method treats the intrinsic SHC
and the extrinsic effects on the same footing via so-called vertex corrections. In the
nonzero-temperature calculations, the contributions due to the vertex corrections are, however, very small compared to the
intrinsic spin Hall conductivity.

The longitudinal charge conductivity determined by the SPR-KKR package refers to bulk. Interface
scattering is known to reduce the conductivity $\sigma_{xx}$ of thin films, which can be estimated
via the Mayadas-Shatzkes (MS) model,\cite{Mayadas1970}
\begin{equation}
\frac{\sigma_{xx}}{\sigma_{xx0}} = \left[1+\frac{3\lambda_\tn{MF}}{8t_\tn{HM}}\left(1+\frac{p}{2}\right) + \frac{3\lambda_\tn{MF}}{2D_\tn{HM}}\left(\frac{r}{1-r}\right)\right]^{-1}.
\end{equation}
Here, $\sigma_{xx0}$ is the bulk conductivity from the SPR-KKR calculation, $t_\mathrm{HM}$ is the
film thickness, $D_\tn{HM}$ is the lateral grain size. For our material system, the electron
mean-free path $\lambda_\tn{MF}$ is calculated as $\lambda_\tn{MF}/\sigma_{xx} = 0.3\times
10^{-15}~\Omega~\tn{m}^2$. A reasonable fit of the data is obtained with
$D_\tn{HM} \approx 5\,\tn{nm}$ and both the specularity parameter $p$ and the grain boundary
reflectivity parameter $r$ set to 0.5.

\begin{figure}[th!]
\includegraphics[width=8.6cm]{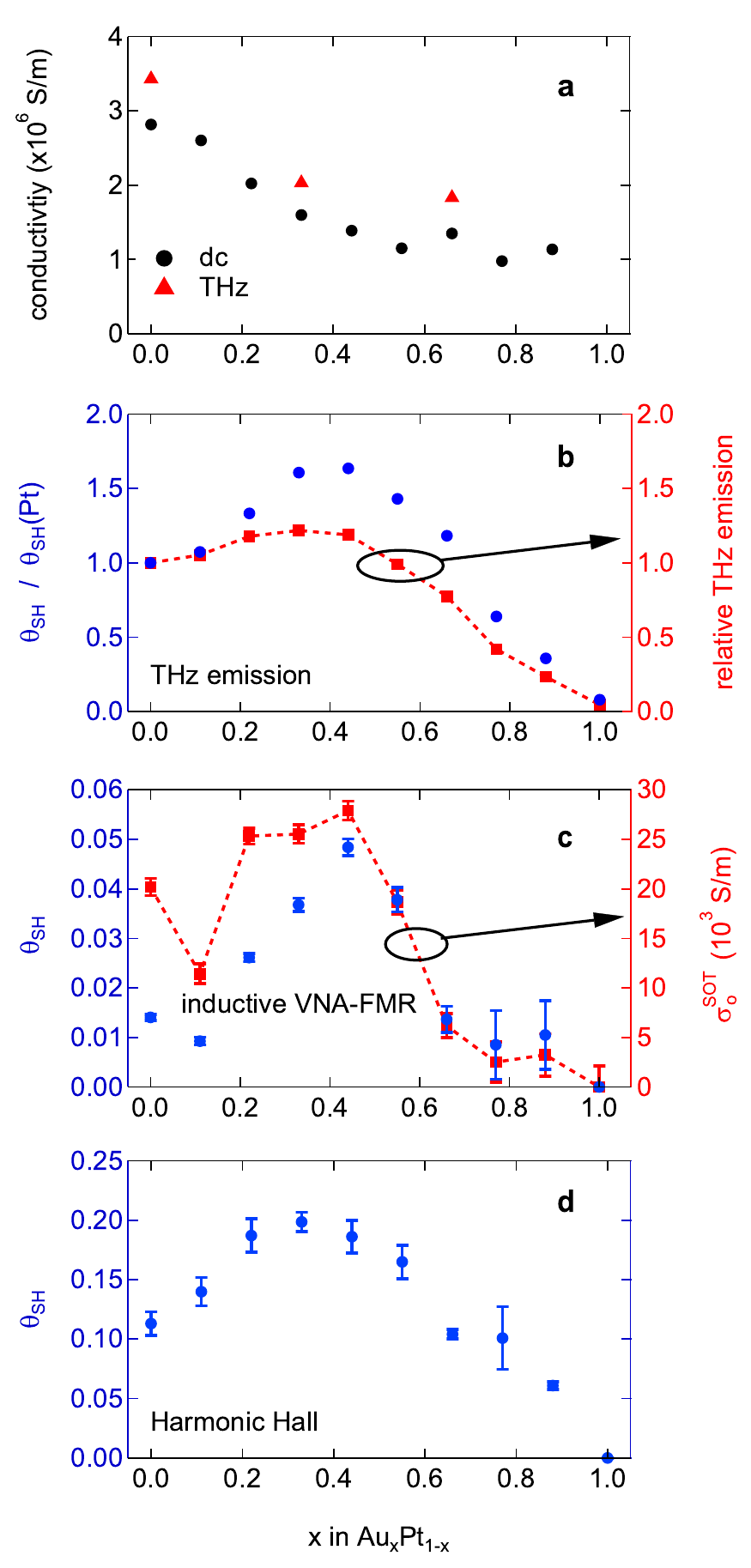}
\caption{\label{spinhallangles}
(a)~Electrical conductivities of the Au$_x$Pt$_{1-x}$ alloy films determined by four-point dc conductivity
 measurements and by THz transmission measurements. In both cases, a parallel conductor model was applied 
to subtract the CoFeB layer conductance. (b)~Relative THz emission (right
axis) and relative spin Hall angle (left axis) as obtained from Eq. \ref{eq:THz}. (c)~Odd component
of the spin-orbit torque conductivity (right axis) obtained in the VNA-FMR measurements and
extracted lower-bound spin Hall angle $\theta_\mathrm{SH} = \sigma_\tn{o}^\tn{SOT} /
\sigma_{xx}$ (left axis). (d) Spin Hall angles as determined with the harmonic Hall response method
via Eq.~\eqref{eq:harmonichall_SHA}.}
\end{figure}

\section*{Results and discussion}

\figus{spinhallangles}b-c show the major results obtained with the three methods employed here.
Regarding TES, \fig{spinhallangles}b displays the THz emission amplitude (right axis) and the SHA
as extracted using \eq{eq:THz} (left axis). Both quantities are normalized to those of the pure Pt
layer ($x=0$). The THz emission
amplitude exhibits a maximum at a Au fraction of $x\approx 0.4$, which is even more pronounced in
the relative SHA. The reason for this difference is the monotonic decay of the HM conductivity
(\fig{spinhallangles}a) and the electron mean-free path leading to a decreasing spin-current
relaxation length $\lambda_\tn{s}$ with increasing $x$.

Concerning the VNA-FMR measurements, \fig{spinhallangles}c shows the odd component of the SOT
conductivity (right axis) and the SHA (left axis) as obtained through \eq{eq:thetaSH_VNAFMR}. The SOT
conductivity features a broad plateau around $x \approx 0.33$ and decays with increasing Au
content. The $\sigma_\tn{o}^\tn{SOT}\approx 2 \times 10^4~\tn{S/m}$  measured for $x=0$ (pure Pt) is in good agreement with $\sigma_\tn{o}^\tn{SOT}\approx 3 \times 10^4~\tn{S/m}$ measured for Pt/NiFe with the same technique~\cite{Berger2018_1}. As the charge conductivity decreases with increasing $x$ (\fig{spinhallangles}a), a local
maximum of $\theta_\mathrm{SH}$ arises around $x\approx 0.4$.

Finally, the SHA as determined by the HHR method is displayed in \fig{spinhallangles}d. These
measurements feature a local maximum around $x \approx 0.33$. As we have determined both the SHA
$\theta_\tn{SH}$ and the conductivity $\sigma_{xx}$ of the HM layer, we can also compute its SHC
through $\sigma_\tn{SH}=\theta_\tn{SH}\sigma_{xx}$.

\begin{figure}
\includegraphics[width=8.6cm]{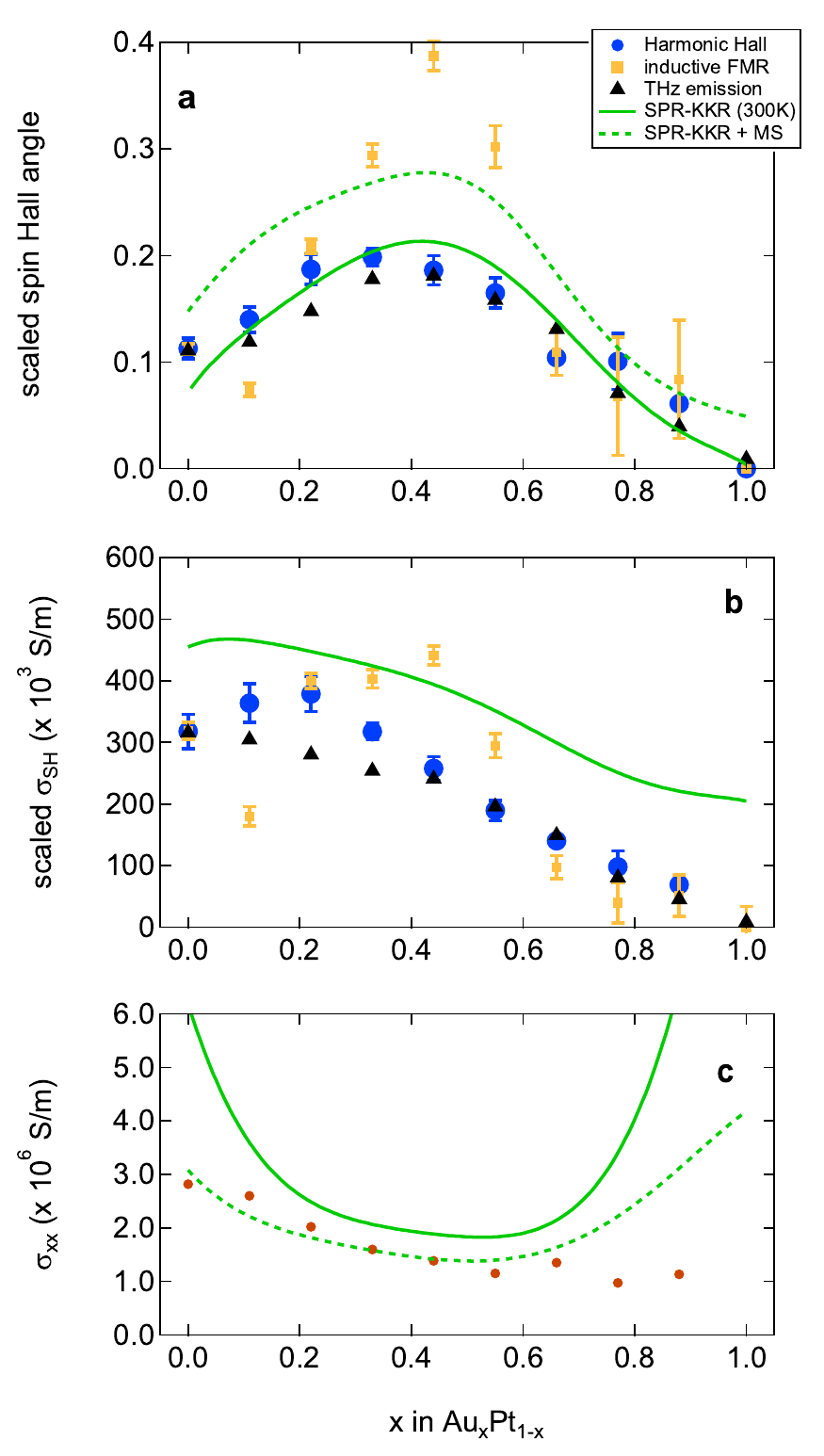}
\caption{\label{scaledresults}
(a) Experimental spin Hall angles scaled to the spin Hall angle of the pure Pt sample obtained with the
harmonic Hall response method. Green lines represent results from SPR-KKR calculations including or
neglecting thin-film corrections of the conductivity via the Mayadas-Shatzkes (MS) model. (b)
Scaled spin Hall conductivities as in (a). The green line represents the SPR-KKR calculation. (c)
Electrical conductivity as measured electrically (dots), as calculated (green line), and as calculated including
corrections from the MS model (dashed line).}
\end{figure}

To better compare the outcome of the three methods, \figs{scaledresults}a and \ref{scaledresults}b
display, respectively, the measured SHAs and SHCs scaled to the HHR results of the pure Pt film
($x=0$). We find that the scaled SHAs vs Au fraction $x$ show similar trends, in particular in
terms of the position of the SHA maximum. All three results are also in reasonable agreement with a
previous experiment on the Au-Pt system.\cite{Obstbaum2016}

In particular, the TES data are in excellent overall agreement with the HHR data with some
discrepancies in both the SHA and SHC around $x \approx 0.2$: The HHR method finds an initial
increase of the SHC with increasing $x$, whereas the TES data suggest a monotonic decrease of the
SHC with increasing $x$. 

While TES delivers SHA and SHC values relative to a reference HM (such as Pt), both VNA-FMR and HHR
provide absolute values. The HHR SHA of the 3~nm pure Pt film ($x=0$) is found to be
$\theta_\tn{SH}^\tn{Pt} \approx 0.11 \pm 0.01$, whereas the maximum SHA at $x = 0.33$ amounts to
$0.20 \pm 0.01$. The $x=0$ value of the SHA agrees very well with other recent measurements on Pt
films with similar conductivities.\cite{Sagasta2016, Nguyen2016, Pai2016} Interestingly, our VNA-FMR SHA
values of pure Pt are a factor of approximately $4$ to $10$ smaller than those from the HHR
measurements. 
Because we have assumed vanishing spin backflow and vanishing SML for evaluation of both VNA-FMR and HHR data [see Eqs.~\eqref{eq:thetaSH_VNAFMR} and~\eqref{eq:harmonichall_SHA}], this discrepancy indicates that at least one of these neglected parameters is substantially different between these techniques. 

Because spin backflow is typically only a small correction for the investigated all-metallic FM/Pt system with effective spin mixing conductances exceeding ~$10^{19}~\tn{m}^{-2}$ for metallic magnets~\cite{Czeschka2011}, we speculate that the discrepancy is predominantly caused by a difference in SML. 

The SML~\cite{Liu2014} has been found to be strong in previous FMR-based experiments~\cite{ Rojas2014, Berger2018_1, Berger2018_2, Keller2019}, where up to 90\% of the spin information can be lost at the metallic FM/Pt interface~\cite{Berger2018_1}. To reconcile our FMR and HHR measurements quantitatively in the context of SML alone, we have to assume a factor 5 to 10 difference in SML between these two techniques. Unfortunately, the SML cannot be unambiguously determined from the existing sample series. We can thus only speculate that the difference might be due to one of the following reasons:
a) The VNA-FMR measurements are performed in out-of-plane geometry, the HHR measurements in the in-plane geometry. 
b) The VNA-FMR measurements are sensitive to the transverse dynamic magnetization components, while the HHR measurements are quasi-static. This could lead to enhanced spin dephasing in the FMR measurements, for instance in a thin layer of proximity-polarized Pt.

Because no absolute values for the SHA can be extracted from the TES measurements, no conclusion about a possible SML at THz frequencies can be drawn. A quantitative evaluation of SML between the different experimental geometries would be highly interesting but is left for future studies that concentrate on a single material system.

In addition to this possible dependence of the SML on experimental geometry, further magnetization-direction dependent corrections to the SHA may exist. For instance, a potential spin-rotation at the interface~\cite{Humphries2017} might depend on the geometry. All these corrections can, in addition, depend on the stoichiometry due to modifications of the film resistivities and the interfacial electronic structure matching. 

To gain more insight into the observed composition dependence of the measured SHA,
\fig{scaledresults} also displays the unscaled first-principles results for the SHA, SHC and the
charge conductivity of Au$_x$Pt$_{1-x}$ vs $x$. The experimental results for the SHA from the HHR
method agree very well with the SPR-KKR calculation without the conductivity reduction from the MS
model (see above). Inclusion of the MS model predicts a larger SHA for all stoichiometries. This
effect can be traced back to the SHC, which is larger in the calculation than in the experiment
(Fig.~\ref{scaledresults}b). At the same time, the film conductivity is smaller than what is predicted
by SPR-KKR including the MS model for large Au content. The large deviation at high Au content can
be attributed to island growth (see above), leading to strong grain-boundary scattering and, thus,
to much lower conductivity than expected from the MS model with a single set of parameters.

Both techniques that provide quantitative results for the SHA and the SHC, the VNA-FMR method and
the harmonic Hall response method, show smaller SHC than predicted by the SPR-KKR calculations.
This trend may be explained by the already mentioned neglect of interfacial spin memory loss. Other
effects that may reduce the spin current include  interfacial spin transparency and interfacial
spin-orbit coupling.\cite{Amin2016} The reduced spin current density manifests itself either as
less damping-like spin-orbit torque observed in the harmonic Hall measurements, or as less detected
charge current density in the inductive VNA-FMR experiment.

\section*{Conclusion}

Although an accurate determination of the internal SHA of a given material is of great fundamental
interest, in high-throughput experiments, it is often sufficient to observe a trend in relative
terms, rather than measuring absolute values. To date, studies on the SHE have mostly focused on
elements and binary alloys and compounds. Therefore, we are only at the beginning of mapping out
the SHE in ternary, quaternary and more complex alloys and compounds. For an
efficient search for materials with large SHA or large SHC, high-throughput techniques for
measuring these properties are necessary, and the two techniques presented here, VNA-FMR and THz
emission spectroscopy, are now proven tools for future experimental work.

\section*{Acknowledgments}
This work was supported by the German research foundation (DFG) through project WE5386/4-1, through
TRR227 \lq\lq Ultrafast spin dynamics" (projects B02 and A05) and by the ERC through ERC CoG TERAMAG (grant
no. 681917).

\section*{Appendix}

\subsection{Sample fabrication} Thin film heterostructures of
Substrate$|$Au$_x$Pt$_{1-x}$(3~nm)$|$Co$_{40}$Fe$_{40}$B$_{20}$(3~nm)$|$Si(1.5~nm) were grown by
magnetron sputtering at room temperature. For the THz emission experiments and for the inductive
GHz measurements, we used polished fused silica substrates. Si wafers with a 50nm thermal oxide
layer were used for the low-frequency harmonic Hall measurements. The Pt-Au alloys were made by
magnetron co-sputtering from two elemental targets. All samples were exposed to a pure oxygen
plasma via the Si source prior to deposition to clean the substrate surface. The Pt-Au
stoichiometries were verified by x-ray fluorescence spectroscopy. For all samples, the film
thicknesses and crystallographic phases were checked by x-ray reflectivity and diffraction
measurements.

For the THz emission spectroscopy and the inductive GHz measurements, the samples were not
subjected to further processing. For the harmonic Hall measurements, the samples on Si$|$SiOx
wafers were lithographically patterned into Hall crosses with an arm width of 16$~\mu\tn{m}$ and
bonded into DIL-24 packages. During the lithographic processing, the films were heated to
90$^\circ$C for 20\,min. No effect of the heating was seen in subsequent conductivity and x-ray
diffraction measurements.

\subsection{THz emission spectroscopy} The THz emission was driven with ultrashort laser pulses from a
Ti:Sa oscillator with pulse duration of 10~fs, central wavelength of 800~nm, repetition rate of
80~MHz, and pulse energy of about 1~nJ. The THz transient was measured via electro-optic
sampling\cite{Leitensdorfer1999} in a 1-mm-thick ZnTe (110) crystal with a weak copropagating 10-fs
near-infrared probe pulse from the same laser.
The electrical conductivities of the AuPt alloy were obtained by THz transmission measurements as 
detailed in Refs. \onlinecite{Seifert2017} and \onlinecite{Seifert2018_2}.

\subsection{VNA-FMR measurements} A static magnetic field $H$ is applied along the bilayer normal. FMR
is excited by passing a microwave current of fixed frequency $\omega$ through the CPW while sweeping
the magnitude of $H$. At each value of $H$, the complex-valued microwave transmission $S_{21}(H)$
through the CPW  is recorded with the VNA. Experiments are repeated for $\SI{10}{\giga\hertz}\leq
\omega/2\pi\leq \SI{40}{\giga\hertz}$. For each $\omega$, the obtained $S_{21}(H)$ spectra are fitted to
\begin{equation}\label{eq:S21fitting}
S_{21}(H)=S_{21}^0 - \mathrm{i} A \frac{\chi(H)}{M_\tn{s}} \;,
\end{equation}
where $S_{21}^0$ is the $H$-independent transmission through the CPW outside FMR conditions, $A$ is
a complex-valued scaling parameter, $\mu_0M_\tn{s}=\SI{1.05}{\tesla}$ is the saturation
magnetization and
\begin{equation}\label{eq:chi}
\chi(H)=\frac{M_\tn{s}\left(H-M_\tn{eff}\right)}{\left(H-M_\tn{eff}+ \mathrm{i} \Delta H\right)^2-H_\tn{eff}^2} \;,
\end{equation}
is the diagonal component of the Polder susceptibility tensor.\cite{Dreher2012} Here,
$M_\tn{eff}=H_\tn{res}-H_\tn{eff}$ and $H_\tn{eff}=\omega/(\mu_0 \gamma)$, where $\gamma$ is the
gyromagnetic ratio. \cite{Nembach2011} After fitting the data to Eq.~\eqref{eq:S21fitting}, as
detailed in \onlinecite{Nembach2011} we extract the normalized inductance \cite{Berger2018_1}
\begin{equation}\label{eq:Ltilde}
\widetilde{L}=\frac{L}{\chi(H_\tn{res})}= \frac{2 A Z_0}{\omega M_\tn{s} S_{21}^0} \;,
\end{equation}
with the impedance of the CPW $Z_0=\SI{50}{\ohm}$. Due to the normalization by $S_{21}^0$ in
Eq.~\eqref{eq:Ltilde}, $\widetilde{L}$ is quantitatively determined without any calibration of the
microwave circuit. Furthermore, the Gilbert damping $\alpha_\tn{tot}$ is obtained by fitting the
$\Delta H$ vs. $\omega$ data to
\begin{equation}\label{eq:alpha}
\Delta H=\frac{\omega}{\mu_0\gamma} \alpha_\tn{tot}+\Delta H_0 \;,
\end{equation}
with the inhomogeneous linewidth broadening $\Delta H_0$. 

\subsection{Harmonic Hall measurements} For the determination of the spin Hall angle, the films were
patterned into 4-fold rotationally symmetric Hall crosses with a conductor width of $w =
16~\mu\tn{m}$ and a length of $l=48~\mu\tn{m}$ by optical lithography. Harmonic Hall voltage
measurements were performed in a dual Halbach cylinder array with a rotating magnetic field up to
1.0~T (MultiMag, Magnetic Solutions Ltd.).  An ac current density with an rms value of $j_\tn{rms}
= 2 \times 10^{10}$~A~m$^{-2}$ ($I_\tn{rms} = 1.92~\tn{mA}$) and frequency $\omega / 2\pi =
3219\,\tn{Hz}$ was injected into the Hall crosses and the in-phase first harmonic and out-of-phase
second harmonic Hall voltages were recorded simultaneously upon in-plane field rotation with a
Zurich Instruments MFLI multi-demodulator lock-in amplifier. The out-of-plane saturation component
of the effective field $B_\tn{eff} = B_\tn{ext} + B_\tn{sat}$ is $B_\tn{sat} = B_\tn{dem} -
B_\tn{ani} > 0$. It was obtained together with the anomalous Hall resistance amplitude $R_A$ from
Hall voltage measurements in a perpendicular magnetic field up to 2.2\,T. The planar Hall
amplitudes $R_\tn{P}$ were obtained from the first harmonic $V_\omega = R_\tn{P} I_\tn{rms} \sin
2\varphi$. The parasitic ANE component $\alpha' I_0$ yields an electric field $\bm{E}_\tn{ANE} =
-\alpha \nabla T \times \bm{m} \propto I_0^2$, where $I_0$ is the current amplitude. The prefactor
$\alpha'$ summarizes all geometrical parameters and the film electrical conductivity, heat
conductivity, etc. that determine $\nabla T$. The magnetization of the CoFeB film was determined by
alternating gradient magnetometry to be $M_\tn{s} = (1050 \pm 50)\,\tn{kA/m}$. The parallel circuit
model was applied to determine the current density flowing in the HM layer.

\subsection{Linear response calculations} The spin Hall conductivities were calculated within a fully
relativistic multiple-scattering Green function framework using the Kubo-Bastin formalism
\cite{Lowitzer2011}. Intrinsic and extrinsic contributions to the spin Hall conductivity are
treated on equal footing. Furthermore, chemical alloying as well as temperature are treated on
equal footing within the coherent potential approximation (CPA), or the alloy-analogy model (AAM),
respectively \cite{Ebert2015}. The formalism is implemented in the Munich Spin-Polarized
Relativistic Korringa-Kohn-Rostoker (SPR-KKR) code \cite{Ebert2011, SPRKKR}. The Green function was
expanded up to $\ell_\tn{max} = 3$ and the Fermi energy was accurately obtained with Lloyd's
formula. The atomic sphere approximation (ASA) was used throughout. Dense k-point meshes were used
to ensure an accurate evaluation of the Brillouin zone integrals for the Fermi surface term. For
more details see Ref. \onlinecite{Obstbaum2016}.




\end{document}